\documentclass[prl,twocolumn,superscriptaddress]{revtex4}

\usepackage{amsmath}

\usepackage{graphicx}
\usepackage{times}

\begin{document}

%%%%%%%%%%%%

\title{Dynamical Disentanglement across a Point Contact in
a Non-Abelian Quantum Hall State}

\author{Paul Fendley}
\affiliation{Department of Physics, University of Virginia,
Charlottesville, VA 22904-4714}
\author{Matthew P.A. Fisher}

\affiliation{Kavli Institute for Theoretical Physics,
University of California, Santa Barbara, CA 93106-4030}
\author{Chetan Nayak}
\affiliation{Microsoft Research, Project Q, Kohn Hall,
University of California, Santa Barbara, CA 93106-4030}
\affiliation{Department of Physics and Astronomy,
University of California, Los Angeles, CA 90095-1547}

\date{April 3, 2006}

\begin{abstract}
We analyze tunneling of non-Abelian quasiparticles
between the edges of a quantum Hall droplet at Landau level filling
fraction $\nu=5/2$, assuming that the electrons
in the first excited Landau level organize themselves
in the non-Abelian Moore-Read Pfaffian state.
We formulate a bosonized theory
of the modes at the two edges of a Hall bar;
an effective spin-$1/2$ degree of freedom emerges in
the description of a point contact. We show how the crossover from
the high-temperature regime of weak quasiparticle tunneling
between the edges of the droplet,
with $4$-terminal $R_{xx}\sim T^{-3/2}$,
to the low-temperature limit, with
$R_{xx} -  \frac{1}{10}\,\frac{h}{e^2} \sim - {T^4}$,
is closely related to the two-channel Kondo effect.
We give a physical interpretation for the entropy of
$\ln(2\sqrt{2})$ which is lost in the flow from
the ultraviolet to the infrared.

\end{abstract}

\maketitle

%%%%%%%%%%%%

\paragraph{Introduction.}
There is indirect evidence \cite{Morf98} that a non-Abelian
topological state may occur at the observed quantized Hall plateau
with $\sigma_{xy}=\frac{5}{2}\,\frac{e^2}{h}$ \cite{Willet87,Xia04}.
The leading candidate is the Moore-Read Pfaffian state
\cite{Moore91,Greiter92}, in which charge-$e/4$ quasiparticles exhibit
non-Abelian braiding statistics
\cite{Nayak96c,Tserkovnyak03,Read96,Fradkin98,Read00,Ivanov01,Stern04}.
This state would support topologically-protected qubits, whose
observation would confirm the non-Abelian nature of this quantized
Hall plateau \cite{DasSarma05}.  Other interference measurements would
also directly probe the non-Abelian braiding statistics of
quasiparticles in this state \cite{Fradkin98,Stern05,Bonderson05,Hou06}.

All of these proposed experiments suggest employing the gapless
edge excitations as a probe of the bulk.  Interedge tunneling of quasiparticles
occurs when two edges of the Hall bar are brought into close proximity.
In these proposals, it is assumed
that tunneling at these contacts is weak.  However,
just as in the Abelian states \cite{Wen91,Kane92,Fendley95}, as the
temperature and voltage are decreased, the effective quasiparticle
tunneling strength increases until the Hall droplet is effectively
split into two at the point contact.  

In this paper, we describe this
crossover precisely.
The non-Abelian statistics of the quasiparticles means that 
care is required to even define the tunneling operator 
at the point contact.  Although
the tunneling occurs at a single point in space, the quasiparticle
which is tunneling can be entangled with others far away.
We find the tunneling operator for the simplest
kind of point contact in the Moore-Read state in terms of the underlying
edge theory, a critical 2d Ising model (the neutral sector) and a free
boson (the charged sector). This allows us
to describe the perturbation expansion of quasiparticle tunneling
processes at the point contact
in terms of the chiral correlators of the edge theory.
We show how to bosonize the tunneling operator, and find the remarkable result
that quasiparticle transport through
the point contact is closely related to the two-channel anisotropic
Kondo effect \cite{Emery92} and resonant tunneling in
Luttinger liquids \cite{Kane92}.

As the droplet breaks in two, the initial entropy, corresponding to
the uncertainty in the non-Abelian topological charges of the two
halves of the droplet, is removed. We show how the difference of
entropies between the system without tunneling (the ultraviolet limit)
and the system split into two droplets (the infrared limit) is
$S_{UV}-S_{IR}=\ln(2\sqrt{2})$.  When only the most relevant tunneling
operator is present, the arrival at the infrared strong-coupling fixed
point is fine-tuned so that the leading irrelevant operator has
scaling dimension $8$ and tunnels a {\it pair} of electrons from one
half of the droplet to the other. When a subleading marginal
tunneling operators is present in the ultraviolet, the leading
irrelevant operator in the infrared has scaling dimension $3$ and
tunnels a single electron.

\paragraph{Edge Excitations at $\nu=5/2$.}

We assume that the lowest Landau level
(of both spins) is filled and the first excited Landau level
is in the Moore-Read state. There will, therefore, be two
integer quantum Hall edge modes, which will be the
outermost excitations of the system. Since we will
be focussing on tunneling across the interior of a Hall droplet,
these modes will participate very weakly. Hence, we
will ignore them and focus on the half-filled
first excited Landau level. The gapless chiral theory
describing the edge excitations has both a charged sector and a
neutral sector. The charged sector is described by a free boson
$\phi_c$.
Edge modes described solely by free bosons are characteristic of abelian
fractional quantum Hall states, and are now very well understood
\cite{Wen91,Kane92,Fendley95}. The
novel properties of the non-Abelian state believed to be realized at
$\sigma_{xy}=\frac{5}{2}\,\frac{e^2}{h}$
arise from the neutral sector, which is described by a critical Ising field
theory \cite{Milovanovic96}.

%which is a free boson, 
%and a neutral sector which is a free neutral, or Majorana, fermion.
%\begin{equation}
%{\cal L}^{\rm edge} =\frac{1}{4 \pi} \partial_x
%\phi_{c}(\partial_t+{v_c}\partial_x)\phi_{c}
%+ \frac{1}{2 \pi}\, \psi(\partial_t+{v_n}\partial_x)\psi
%\label{eqn:edge-Lagrangian}
%\end{equation}
%In the first term in (\ref{eqn:edge-Lagrangian}),
%the charged boson has compactification radius 
%The velocities of the charge and spin modes are, in general,
%different, and we expect ${v_c}>{v_n}$.

The chiral part of the $1+1$ dimensional Ising field theory 
contains a spin field $\sigma$ and a free Majorana fermion $\psi$.
Correlators involving only the fermion are trivial to
compute, but finding those involving the Ising spin field requires
much more work.  One reason the physics of the Moore-Read state is so
interesting is because the operator which creates or annihilates a charge $e/4$
quasiparticle contains the chiral part of the spin field. The
spin field is not local with respect to the Majorana fermion: it creates
a branch cut for the fermion field. When a charge $e/4$ quasiparticle
tunnels to or from an edge of a droplet, the fermion boundary conditions
around the Hall droplet change: they are anti-periodic
when there is an even number of $e/4$ quasiparticles
in the bulk, and periodic when there is an odd number.

\begin{figure}[t!]

\centerline{\includegraphics[width=3.45in]{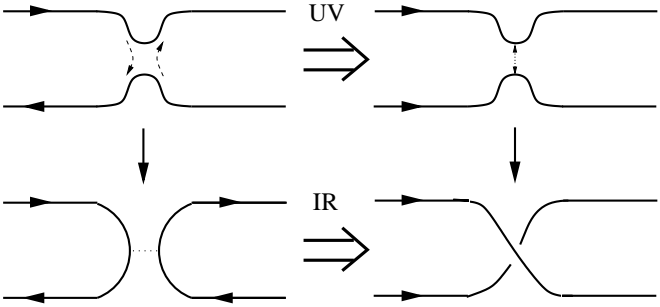}}

\caption{We ignore the $\nu=2$ edge excitations of the filled
lowest Landau level (of both spins) and focus on the
$\nu=1/2$ edge excitations. To compute their correlation
functions, we redraw the lower edge so that its excitations
also propagate to the right. This allows us to bosonize the
neutral fermionic excitations of both edges.}

\label{fig:flipped}

\end{figure}

We study quasiparticle tunneling between two different points on the
edge of a Hall droplet. The quasiparticle of charge $e/4$ is created
by the operator $\sigma e^{i\phi_c/(2\sqrt{2})}$, where the chiral
boson $\phi_c$ is normalized so that $e^{ia\phi_c}$ has dimension
$a^2/2$. A chiral Ising spin field has dimension $1/16$, so the operator
describing tunneling of one such quasiparticle from one edge to the
other has dimension $1/4$.  Other tunneling processes are described by
the dimension-1/2 operator tunneling the charge-$e/2$ quasiparticle
created by $e^{i\phi_c/\sqrt{2}}$, and the marginal operator tunneling
the neutral fermionic quasiparticle $\psi$. The
tunneling Lagrangian is schematically
\begin{multline}
{\cal L}^{\rm tun} \: = \: \lambda_{1/4} \, {\sigma_\alpha}
{\sigma_\beta} \,e^{i (\phi_{c\alpha} - \phi_{c\beta})/2\sqrt{2}} \: + \: \text{h.c.}\\
+ \: \lambda_{1/2} \, e^{i (\phi_{c\alpha} - \phi_{c\beta})/\sqrt{2}} + \text{h.c.} \:
+ \: i \lambda_{1} {\psi_\alpha} {\psi_\beta}
\label{eqn:tunnel-op-Lagrangian}
\end{multline}
where the subscripts $\alpha,\beta$ refer to the spatial points on
either side of the point contact, as in Fig.  \ref{fig:flipped}.  We
say ``schematically'' because as we will describe below, the operator
$\sigma_\alpha\sigma_\beta$ is non-local and requires additional
information to be defined precisely. In the limit that $\lambda_{1/4}$
is small, we can ignore such complications to compute the
leading-order behavior \cite{Bena06}. We expect $\lambda_{1/2} \ll
\lambda_{1/4}$ (because $\lambda_{1/2} \sim \lambda_{1/4}^2$), in
which case the $4$-terminal longitudinal resistance (defined
as the voltage drop along one edge of the Hall bar
divided by the transmitted current) scales as
\begin{equation}
R_{xx} \sim \lambda_{1/4}^2\,T^{-3/2} .
\label{eqn:pert-sigma-xx}
\end{equation}
At finite voltage $V>T$, we instead have $I \sim V^{-1/2}$.

\paragraph{Chiral Correlation Functions.}

Since the tunneling amplitudes $\lambda_{1/4}$ and $\lambda_{1/2}$
increase in importance as the temperature decreases
(\ref{eqn:pert-sigma-xx}), the weak-tunneling regime does not extend
to arbitrarily low temperatures.  Therefore, we must go beyond the
lowest-order perturbation theory contribution given in
(\ref{eqn:pert-sigma-xx}).  This requires understanding correlators of
chiral Ising fields; more information than just the space-time
locations of the fields must be provided in order to define
correlators uniquely.  For example, the four-point function of
non-chiral Ising spin fields does not decompose into a simple product
of a chiral correlator times its antichiral conjugate, but rather is a
sum of such terms \cite{Belavin84}.  In other words, local Ising
fields cannot simply be decomposed into a product of a left-moving
part times a right-moving part.  In the Moore-Read state, this is a
manifestation of the non-Abelian statistics of the quasiparticles.

To understand the situation in more depth, note that a
pair of quasiparticles can be in either of two topologically
distinct states \cite{Moore91,Nayak96c,Read96,Read00},
the two states of the qubit which they form \cite{DasSarma05}.
A simple physical picture arises from considering
the closely-related system of a superconductor with pairing symmetry
$p+ip$. Charge-$e/4$ quasiparticles
correspond to half-flux quantum vortices in this superconductor;
there is a Majorana fermion zero mode
in the core of each such vortex \cite{Read00}.
The Majorana fermion zero modes
associated with two vortices can be combined into a Dirac fermion
zero mode, which can be either occupied or unoccupied -- the two
states $|0\rangle$ and $|1\rangle$ of the qubit.
In this way \cite{Read00}, one can show that there are
$2^{n-1}$ states of $2n$ quasiparticles \cite{Nayak96c,Read96}.
When quasiparticles are braided, these states are rotated into each
other according to the spinor representation of $SO(2n)$ \cite{Nayak96c,Ivanov01}.
%(As far as non-Abelian braiding properties are concerned,
%quasiparticles and quasiholes are identical, and we will sometimes use
%`quasiparticles' to refer to both of them.)

The same issue arises in a correlation function of $2n$
quasiparticle operators.  There is a $2^{n-1}$-dimensional
vector space of such correlation functions -- called {\it conformal
blocks} \cite{Belavin84} -- which transform into each other as the
quasiparticle positions are taken around each other in
$1+1$-dimensional spacetime.  This is a consequence of the fact that 
the operator product of two spin fields
contains both the identity field and the fermion; 
this ``fusion'' is written schematically as $\sigma \cdot\sigma \sim
I + \psi$.  Combining the two languages, we see that when a pair of
quasiparticles is in the state $|0\rangle$, they fuse to $I$; when
they are in the state $|1\rangle$, they fuse to $\psi$.

To define precisely the tunneling operator and its conformal blocks,
we must therefore also specify the state of the qubit associated with
the quasiparticle which tunnels and the quasihole which it leaves
behind. We assume that tunneling processes do not create additional
neutral fermions in the bulk because the topological state of the
system cannot be affected by a tiny motion (tunneling from one side of
the point contact to the other) of a quasiparticle.  This means that
when a quasiparticle tunnels from one edge to another, it forms a
qubit in state $|0\rangle$ with the quasihole which is left behind.
%Correlation
%functions which arise in the perturbative
%expansion of (\ref{eqn:tunnel-op-Lagrangian})
%are then uniquely specified.  
This assumption fixes uniquely the 
conformal blocks occuring in the perturbative expansion of
(\ref{eqn:tunnel-op-Lagrangian}) in power of $\lambda_{1/4}$.  
The term with $n$ tunneling operators is an element of a $2^{n-1}$
dimensional vector space of conformal blocks. We label the basis elements of
this vector space by $[{m_1}, {m_2},
..., {m_n}]$ with ${m_i} = 0$ or $1$. This means that the $i$th pair
${\sigma_\alpha} {\sigma_\beta}$ fuses to $I$ if ${m_i}=0$ and to
$\psi$ if ${m_i}=1$.  
%Since the conformal block is only non-zero
%if ${\sum_i}{m_i} \equiv 0\,(\text{mod}\,2)$, for four spin fields
%there are only $2$ linearly independent conformal blocks: $[0,0]$ and
%$[1,1]$.
Our assumption means that a formal expression such as $\left\langle
{\sigma_\alpha} {\sigma_\beta}\, {\sigma_\alpha} {\sigma_\beta} \ldots
\right\rangle$ arising in the perturbative expansion is given by
the conformal block $[0,0,\ldots]$.

\paragraph{Cluster decomposition.}

Provided the measurement time scale is short compared to the time for
an excitation to propagate around the droplet from one
side of the contact to the other, 
%one can study the physics of the
%point contact by considering the droplet to be arbitrarily long.
%In this limit, 
one can treat the
two sides of the point contact as two independent edges (so
$\sigma_\alpha$ and $\sigma_\beta$ as well as $\phi_{c\alpha}$ and
$\phi_{c\beta}$ are independent fields). One can then use
cluster decomposition to break each correlator into a product of
correlators in two separate models, e.g.\
\begin{equation}
\left\langle {\sigma_\alpha} {\sigma_\beta}\,
{\sigma_\alpha} {\sigma_\beta} \ldots
\right\rangle \sim
\left\langle {\sigma_\alpha}\, {\sigma_\alpha} \ldots \right\rangle_\alpha\:
\left\langle {\sigma_\beta}\, {\sigma_\beta} \ldots \right\rangle_\beta
\label{eqn:cluster}
\end{equation}
The complication is that the left-hand-side of (\ref{eqn:cluster}) is
specified by how each $\sigma_\alpha(\tau_a)$ fuses with
$\sigma_\beta(\tau_a)$  
%(with $\tau_j,\tau_j^\prime$ 
(with $\tau_a$ the imaginary time of the $a$th tunneling event), but
we would like to treat $\sigma_\alpha$ and
$\sigma_\beta$ on the right-hand-side as independent fields. We thus
need to disentangle the fields $\sigma_\alpha$ and $\sigma_\beta$ in
order to break the left-hand-side into the product of two independent
correlators, and then specify how neighboring
$\sigma_\alpha$'s fuse with each other in $\left\langle
{\sigma_\alpha}\, {\sigma_\alpha} \ldots 
\right\rangle_\alpha$ and neighboring $\sigma_\beta$'s  fuse in $\left\langle {\sigma_\beta}\, {\sigma_\beta} \ldots
 \right\rangle_\beta$.

To overcome these complications, we utilize the work of Moore and
Seiberg, where relations among conformal blocks are derived by utilizing a
variety of consistency conditions \cite{Moore88}.  Because of the
non-abelian structure, braiding (exchanging the order of operators)
not only results in phases, but can change the fusion channels as
well.  In this way we can change from the basis in which each
$\sigma_\alpha(\tau_a)$ fuses in a definite way with its neighboring
$\sigma_\beta(\tau_a)$,
% as on the left-hand side of Eq.~(\ref{eqn:cluster}), 
to a basis in which each $\sigma_\alpha(\tau_{2j-1})$
fuse in a definite way with $\sigma_\alpha(\tau_{2j})$, and similarly for the
$\sigma_\beta$'s. Letting $\mu=\alpha$ or $\beta$, 
we label this new basis by $(m_1,m_2,\dots,
m_{n/2})_\mu$ where $m_j=0,1$ when $\sigma_\mu(\tau_{2j-1})$ and
$\sigma_\mu(\tau_{2j})$ fuse to be into the $I$ and $\psi$ channels. 
We find \cite{long-version}
\begin{multline}
[0,0,\dots ,0] = 
\sum_{\{m_j\} = 0,1} 
\prod_{\mu=\alpha,\beta} \left({m_1},{m_2},\ldots,{m_{n/2}}\right)_\mu  .
\label{eqn:new-basis}
\end{multline}
%which is non-zero only when ${\sum_k}{m_k}$ even. 
With (\ref{eqn:tunnel-op-Lagrangian})
and (\ref{eqn:new-basis}), we now have unambiguous
expressions for the dynamics of the point contact, treating the two sides as independent
edges. The conformal blocks
$\left({m_1},{m_2},\ldots \right)_{\mu}$ can be computed by solving
differential equations \cite{Belavin84}. However, the results are
quite complicated, and it is difficult to extract much intuition from
them. Therefore, it is useful to express them in a different fashion.

\paragraph{Bosonized Formulation.}
Typically, one treats $\psi_\alpha$ and $\psi_\beta$
as the right- and left-handed
parts of a single non-chiral theory. However,
it is more convenient here to treat them as having the same
chirality by flipping the chirality of one edge, as in Fig. \ref{fig:flipped}.
Then, the two chiral Majorana fermion fields form a
single chiral Dirac fermion field, which can be bosonized,
${\psi_\alpha}+i{\psi_\beta} \sim \,e^{i\phi_\sigma}$.  Here the
chiral bosonic field $\phi_\sigma$ is normalized
so that $e^{ia\phi_\sigma}$ has dimension $a^2/2$.
With this bosonization scheme
a semiclassical (instanton) analysis is possible for the
the point contact dynamics in the IR limit. Moreover, it enables us to exploit
similarities with the problem of resonant tunneling between
Luttinger liquids \cite{Kane92}, the Kondo problem \cite{Emery92},
and dissipative quantum mechanics \cite{Leggett87}.

To bosonize our chiral correlators of spin fields
we use the methods of Ref. \onlinecite{Moore88}. The result is:
\cite{long-version}
\begin{multline}
\hskip-10pt
\prod_{\mu=\alpha,\beta} \left({m_1},{m_2},\ldots,{m_{n/2}}\right)_\mu =
\Bigl\langle \prod_{j=1}^{n/2} 
%{\cal A}(m_j) \rangle$,
\Big(e^{i\left({\phi_\sigma}({\tau^{}_{2j-1}})
-{\phi_\sigma}({\tau^{}_{2j}})\right)/2}\\
+(-1)^{m_j}
e^{-i\left({\phi_\sigma}({\tau^{}_{2j-1}})
-{\phi_\sigma}({\tau^{}_{2j}})\right)/2} \Big)\Bigr\rangle
\end{multline}
Performing the sums in
(\ref{eqn:new-basis}) gives the nice result:
\begin{equation}
[0,0,\ldots,0]= 
 \Bigl\langle \prod_{j=1}^{n/2}
 e^{i\left({\phi_\sigma}({\tau^{}_{2j-1}})-
{\phi_\sigma}({\tau^{}_{2j}})\right)/2}  \Bigl\rangle.
 \end{equation}
Remarkably, the right-hand-side is the same as
$\left\langle\left({S^+}e^{-i{\phi_\sigma}/2}+{S^-}e^{i{\phi_\sigma}/2}\right)
\left({S^+}e^{-i{\phi_\sigma}/2}+{S^-}e^{i{\phi_\sigma}/2}\right)
\ldots \right\rangle $ where $\vec{S}$ is a single spin-$1/2$ degree
of freedom. Consequently, the perturbation expansion of ${\cal L}^{\rm
tun}$ in eqn.\ (\ref{eqn:tunnel-op-Lagrangian}) is identical to that of
\begin{multline}
\label{eqn:Kondoesque}
\tilde{\cal L}^{\rm tun} = \lambda_{1/4} \left({S^+}e^{-i{\phi_\sigma}/2}+
{S^-}e^{i{\phi_\sigma}/2}\right) \cos\!\left({\phi_\rho}/2\right)\\
+ \: \lambda_{1/2} \cos{\phi_\rho} \: + \: \lambda_{1}\, {\partial_x}{\phi_\sigma} ,
\end{multline}
with a charge boson
${\phi_\rho}\equiv \left(\phi_{c\alpha} - \phi_{c\beta}\right)/\sqrt{2}$.
Thus, in translating chiral correlation functions of quasiparticle
tunneling operators into bosonic language, we see
the emergence of an effective spin-$1/2$ degree of freedom.
This shows that the dynamics of a point contact in a Moore-Read
non-Abelian quantum Hall state is 
a variant of the two-channel Kondo problem and equivalent
to resonant tunneling between two $g=2$ Luttinger
liquids.  Upon setting $\phi_\rho=0$ (appropriate for
a $p+ip$ superconductor), the neutral sector 
is literally the single-channel (anisotropic) Kondo Hamiltonian.

\paragraph{Kondo Crossover.}
Consider, first, the case in which only the most relevant
tunneling amplitude, $\lambda_{1/4}$, is non-zero. 
In this case the Hamiltonian in (\ref{eqn:Kondoesque})
is invariant under ${\phi_\sigma}\rightarrow-{\phi_\sigma}$
together with a $\pi$ rotation of the spin about the $x-$axis,
a Kramers-Wannier duality symmetry for the non-chiral Ising model.
Under the replacement  $\phi_{\rho,\sigma} \rightarrow 2 \phi_{\rho,\sigma}$,
the tunneling term is identical to the two-channel Kondo
model with ${J_z}=0$, $J_{x,y}=\lambda_{1/4}$.
By performing a unitary transformation, 
$U=\exp(i{S^z}{\phi_\sigma}/2)$, the tunneling Hamiltonian
becomes:
\begin{equation}
U \,{\cal H}_{\rm tun}\, {U^\dagger} =
\lambda_{1/4} {S^x} \cos\!\left({\phi_\rho}/2\right)
+ \pi {v_n} {S^z}{\partial_x}{\phi_\sigma} .
\label{eqn:Toulouse}
\end{equation}
The Toulouse \cite{Emery92} limit corresponds to
dropping the final term.  Although the presence of this marginal
perturbation does change the
dimensions of the operators in the UV, it does not greatly 
effect physical properties in the IR. To see this, note that $\lambda_{1/4}$ is 
strongly relevant and grows in the IR, so that the energy is minimized with  
${S^x}=+1/2$, ${\phi_\rho}=2(2n+1)\pi$ or ${S^x}=-1/2$,
${\phi_\rho}=2(2n)\pi$.  In this limit, the charge mode is
completely reflected, so that $R_{xx} =
\frac{1}{10}\,\frac{h}{e^2}$. (This peculiar value is the voltage
drop due to the complete backscattering of the $\nu=1/2$ edge
divided by the transmitted current carried by the $\nu=2$ edges.)
%From the canonical transformation
%generated by $U$ above, we see that the neutral boson, $\phi_\sigma$,
%receives a phase shift of $\pm \pi/2$, which corresponds in the
%fermionic language to ${\psi_1}\rightarrow \pm{\psi_2},
%{\psi_1}\rightarrow \mp{\psi_2}$.  
At any of these minima,
$\langle{S^z}\rangle=0$, and fluctuations in ${S^z}$ can be integrated
out, generating terms such as $({\partial_x}{\phi_\sigma})^2 \sim
{\psi_1}{\partial_x}{\psi_1} + {\psi_2}{\partial_x}{\psi_2}$ (which
does not couple the two edges) and $({\partial_x}{\phi_\sigma})^4 \sim
{\psi_1}{\partial_x}{\psi_1} \,{\psi_2}{\partial_x}{\psi_2}$, which
couples the energies of the fermionic modes at the two edges.

Irrelevant perturbations at
the infrared fixed point correspond to instantons 
connecting the minima of (\ref{eqn:Toulouse}). 
The instanton
$\Delta{\phi_\rho}$=$\pm 4\pi$, $\Delta{\phi_\sigma}$=$0$,
$\Delta {\bf S}$=$0$ corresponds to the operator,
\begin{equation}
H^{\rm tun}_{\rm pair} = {v_2} \, \cos(4{\phi_\rho}) ,
\end{equation}
which tunnels a charge-$2$ boson between the two droplets.
When $\lambda_1=0$ in the UV, duality implies that this
dimension-$8$ operator is the leading irrelevant
tunneling operator in the IR, giving $R_{xx} -  \frac{1}{10}\,\frac{h}{e^2}
\sim - {v_2^2}\,T^{14}$ as $T\rightarrow 0$.

With $\lambda_1 \ne 0$, the tunneling Hamiltonian 
in the UV is no longer invariant under Kramers duality,
$\phi_\sigma \rightarrow -\phi_\sigma$.
The operator that tunnels electrons between the two drops
in the IR,
\begin{equation}
\label{eqn:electron-tunneling}
H^{\rm tun}_{\rm el} = {v_1} \,{\partial_x}{\phi_\sigma}\,
\cos(2{\phi_\rho}) = {v_1}\, i{\psi_1}{\psi_2}\, 
\cos(2{\phi_\rho}) ,
\end{equation}
is then not forbidden by symmetry.
In this generic case,
the leading low-temperature correction to complete
backscattering at the contact is
determined by this dimension-$3$ electron tunneling operator:
$R_{xx} -  \frac{1}{10}\,\frac{h}{e^2} \sim - {v_1^2}\,T^{4}$.
%\begin{equation}
%\sigma_{xx} -  \frac{1}{2}\,\frac{e^2}{h} \sim - {v_1^2}\,{T^4}
%\end{equation}

\paragraph{Entropy Loss.}
The two states of the spin one-half degree of freedom that emerges in the
weak-tunneling limit, ${S^z}=\pm 1/2$, correspond physically to
whether there is an even or an odd number of bulk charge-$e/4$
quasiparticles to the left (say) of the point contact.  In the infrared limit this spin is screened. Therefore, ground state entropy is lost in the
flow.  In the absence of the charged mode (as for a $p+ip$ superconductor),
the model is equivalent to the single-channel Kondo
problem. The entropy loss here is therefore simply
$\ln 2$, since a spin-$1/2$ degree of freedom in the UV is completely
screened in the IR. With the charged mode present, the entropy loss is larger.
This follows from the
Toulouse limit of Eq.~($\ref{eqn:Toulouse}$).
With $v_n=0$, one can set $S^x=1$, and the problem reduces to a boundary sine-Gordon model
(a pure $\cos(a\phi)$ boundary perturbation). The entropy loss for a
boundary sine-Gordon tunneling operator of dimension $a^2/2$ is
$-\ln(a^2/2)/2$ \cite{Fendley94}. Therefore the entropy loss in our full
problem is $\ln(2\sqrt{2})$.

This entropy loss is in accord with a general result coming from
conformal field theory \cite{Cardy89,Affleck91}. The entropy of a Hall
droplet with trivial total topological charge and perimeter $L$ is $S=
\alpha L - \ln{\cal D}$, where $\alpha=\pi c\, T/12$, $T$ is the
temperature, $c$ is the central charge of the conformal field
theory describing the edge modes, and ${\cal D}$ is the total quantum
dimension of the particular topological state of matter
\cite{Kitaev05,Levin05}. When a droplet breaks into two droplets, each
of which has trivial topological charge, the entropy of the two droplets is $S=
\alpha{L_1} + \alpha{L_2} - 2\ln{\cal D}$, so that the loss in entropy
upon breaking is $\ln{\cal D}$.
The edge theory for the
Moore-Read state is the Neveu-Schwarz sector of the
second $N=2$ supersymmetric minimal model \cite{Milovanovic96},
which has central charge $c=3/2$, and quantum dimension ${\cal D}=2\sqrt{2}$.
For the $p+ip$ superconductor, the edge theory is the Ising model,
which has $c=1/2$ and ${\cal D}=2$. In both cases, the entropy loss at
the point contact is the same as that we deduced above.
Physically, the 
decrease in
entropy arises because
there is no longer any uncertainty in the topological charge of the
two sub-droplets once they break apart.  
Remarkably, the $T=0$ entanglement entropy determined from the reduced density matrix
of a region of perimeter $L$ inside a much larger Hall droplet 
is also of the form,  $S= \alpha L - \ln{\cal
D}$, with $\alpha$ non-universal\cite{Kitaev05,Levin05}.
This suggests that the
entropy loss at the point contact is actually a topological
entanglement entropy. 

\paragraph{Acknowledgements}
We would like to thank M.\ Freedman, E.-A.\ Kim, A.\ Kitaev,
A.\ Ludwig, J.\ Preskill, N.\ Read, and A.\ Stern for discussions.
This research has been supported by the NSF under grants 
DMR-0412956 (P.F.), PHY-9907949 and DMR-0529399 (M.P.A.F.) and
DMR-0411800 (C.N.), and by the ARO under grant W911NF-04-1-0236 (C.N.).

\vskip -0.5cm

%\bibliography{../corr}
%\bibliographystyle{prsty}

\end{document}